\documentclass[conference]{IEEEtran}
\IEEEoverridecommandlockouts

\usepackage{fancyhdr}
\usepackage{url,cite}
\usepackage{amsmath,amssymb,amsfonts}
\usepackage{algorithmic}
\usepackage{graphicx}
\usepackage{textcomp}
\usepackage[dvipsnames]{xcolor}
\usepackage{hyperref}
\hypersetup{
    colorlinks=true,
    linkcolor=Blue,
    urlcolor=Blue,
    citecolor=BrickRed
}
\def\BibTeX{{\rm B\kern-.05em{\sc i\kern-.025em b}\kern-.08em
    T\kern-.1667em\lower.7ex\hbox{E}\kern-.125emX}}

\fancypagestyle{FirstPage}{
\lfoot{\footnotesize \copyright 2024 Pindrop Security, Inc. This work is licensed under \href{https://creativecommons.org/licenses/by-nc-nd/4.0/legalcode.en}{CC BY-NC-ND 4.0}.\hfill} 
}

\begin{document}

\title{Phonetic Richness for Improved Automatic \\ Speaker Verification}

\author{\IEEEauthorblockN{Nicholas Klein}
\IEEEauthorblockA{
\textit{Pindrop}\\
Atlanta, GA, USA \\
nklein@pindrop.com}
\and
\IEEEauthorblockN{Ganesh Sivaraman}
\IEEEauthorblockA{
\textit{Pindrop}\\
Atlanta, GA, USA \\
gsivaraman@pindrop.com}
\and
\IEEEauthorblockN{Elie Khoury}
\IEEEauthorblockA{
\textit{Pindrop}\\
Atlanta, GA, USA \\
ekhoury@pindrop.com}
}

\maketitle

\begin{abstract}
    When it comes to authentication in speaker verification systems, not all utterances are created equal. It is essential to estimate the quality of test utterances in order to account for varying acoustic conditions. In addition to the net-speech duration of an utterance, it is observed in this paper that phonetic richness is also a key indicator of utterance quality, playing a significant role in accurate speaker verification. Several phonetic histogram based formulations of phonetic richness are explored using transcripts obtained from an automatic speaker recognition system. The proposed phonetic richness measure is found to be positively correlated with voice authentication scores across evaluation benchmarks. Additionally, the proposed measure in combination with net speech helps in calibrating the speaker verification scores, obtaining a relative EER improvement of 5.8\% on the Voxceleb1 evaluation protocol. The proposed phonetic richness based calibration provides higher benefit for short utterances with repeated words.
\end{abstract}


\section{Introduction}
\label{sec:intro}
\thispagestyle{FirstPage}

Automatic Speaker Verification (ASV) systems are increasingly used to authenticate users based on their voice for various secure transactions. 
There is an increasing user-experience driven demand for securely authenticating with shorter utterances of free flowing speech.
Short utterances are those which contain 1 to 8 seconds of net-speech~\cite{Zeinali2019}. 
Net-speech is the duration of actual speech content within a given audio as determined by the speech activity detection.
When the utterances are shorter, the accuracy of ASV varies greatly based on the amount of net-speech and the signal-to-noise ratio (SNR)~\cite{mandasari2015quality}.

However, in real-world applications where the ASV system is passively used and no fixed phrases are required, it is possible for the speaker to choose to repeat the same word or expression several times.
For example, the repetition of the words ``agent" or ``representative" is very common in call center applications where some customers try to avoid talking to the automated interactive voice response (IVR) system, and want to speak to a real agent.
Thus, utterances with repeated words may satisfy the minimum net-speech requirements of an ASV system, but due to their low phonetic diversity, the utterances would still be equivalent to a low quality utterance. 
It is therefore essential to estimate the quality of the test audio to improve the authentication of short utterances.

\begin{figure}[t]
    \centering
    \includegraphics[width=\linewidth]{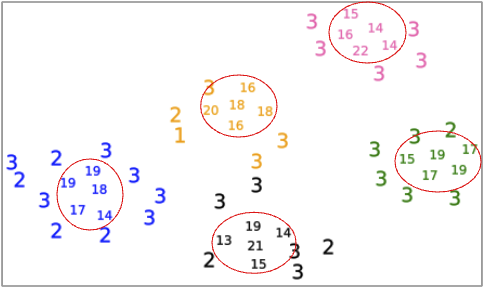}
    
    \caption{\small{T-SNE plot of ECAPA-TDNN speaker verification embeddings for utterances with similar net-speech but varying number of unique phonemes, spoken by APLAWD speakers `a' through `e'. Number markers indicate the number of unique phonemes in the utterance while their color distinguishes speakers. Clusters of utterances with more phonemes are circled in red.}}
    \label{fig:vb_tsne_3} 
\end{figure}

Quality-aware calibration has been explored in various works where a quality measure function (QMF) is used as side information for the calibration model to adjust the raw score. 
This enables the calibration model to adapt to varying conditions in the enrollment and test utterances. 
Various QMFs have been explored in the literature, including: the utterance embedding magnitude~\cite{Thienpondt2020TheIV}; the expected mean imposter score~\cite{Thienpondt2020TheIV}, a measure related to the popular adaptive symmetric normalization (as-norm)~\cite{karam2011towards} method; noise features such as SNR~\cite{Tan2018DNNBasedSC,rao2021improving}; and duration-related features such as duration or net speech, where both the duration of the enrollment and the test utterance can be leveraged~\cite{Thienpondt2020TheIV, Lavrentyeva2022InvestigationOD, Cumani2021AGM, Mandasari2013QualityMF, 6639154,rao2021improving}. 

Duration QMFs are of particular interest in short utterance speaker verification as the performance of ASV systems is known to degrade when there is an insufficient quantity of speech on either the enrollment or test side~\cite{Poddar2017SpeakerVW}. 
There is a consensus amongst these works that the short duration is challenging due to the higher variation in the lexical content leading to higher variation of the speaker embeddings belonging to the same speaker~\cite{larcher:hal-01927726, KANAGASUNDARAM201469}.
For example, a speaker simply blabbering ``blah-blah", coughing, or clearing their throat would be considered as poor quality utterances as they do not contain enough phonetic content normally seen in continuous speech.

In Fig.~\ref{fig:vb_tsne_3}, we observe that utterances with more unique phonemes appear to be clustered more tightly, while those with fewer unique phonemes are more spread out despite all visualized utterances in the plot having similar net-speech.

Authors in~\cite{KANAGASUNDARAM201469} experimented with addressing this issue at the i-vector level through short-utterance-variance-normalization and~\cite{6639154} applied duration-based score calibration.
However, neither of these works have explicitly focused on the issue of phonetic variance in short utterances directly, but rather utilized duration as a proxy.

In this paper, we explore the importance of the diversity of phonetic content for speaker verification.
Particularly, we define phonetic richness as the number of unique phonemes present in a given utterance.
The phonetic richness measure is positively correlated to the raw cosine score between the enrollment and test embedding.
In order to explore the unequal contribution of phonemes towards speaker verification score, we define a phoneme-weighted phonetic richness measure and learn the phonetic weights from data.
Section~\ref{sec:formulations} describes the proposed phonetic richness measures.
We perform score calibration using the phonetic richness measures on short-utterance speaker verification protocols. 
The details of the datasets and evaluation protocols used are provided in~\ref{sec:datasets} and that of the speaker verification system is given in~\ref{sec:system}.
The experiments and results presented in~\ref{sec:eval} show that the proposed phonetic richness measures provide improved score calibration performance over net-speech and are also complimentary to net-speech as a quality measure for score calibration.

\section{Phonetic richness measures}
\label{sec:formulations}
The phonetic richness of an utterance can be quantified using the orthographic transcription and a grapheme to phoneme conversion~\cite{yolchuyeva2019grapheme,g2pE2019} system. 
We define phonetic richness in terms of the number of unique phonemes present in the utterance, consistent with the phonetic measure explored in~\cite{6639154}. 
We obtain automatic phonetic transcription of an utterance using 39 phonetic alphabets of the English language. 
For a given phonetic transcription of an utterance, we can create an N-dimensional binary vector $\boldsymbol{p_u}$ of phoneme presence, where $p_{u,i}, (i=1..N)$ is the binary flag indicating the presence of the phoneme $P_i$ in the transcription. 
Count-unique ($CU$) is thus defined as
\begin{equation}
    CU = \sum_{i=1}^{N} p_{u,i}
\end{equation}\label{eq1}

In order to explore whether some phonemes are more important than others for speaker verification, we define the weighted count-unique (WCU) measure which aggregates a weighted sum of phoneme presence.
Formally, WCU for a given utterance $u$ is defined as 
 \begin{equation}
     WCU = \boldsymbol{w} \cdot \boldsymbol{p_u}
 \end{equation}\label{eq2}
where $\boldsymbol{w}$ is a vector of positive real-valued weights and $\boldsymbol{p_u}$ is a binary vector of phoneme presences for a given utterance. $\boldsymbol{w}$ is learned by fitting a linear regression model to predict the speaker-match score $\boldsymbol{w} \cdot \boldsymbol{p_u} = score_u$ using least squares regression where $score_u$ is a score that a given speaker verification system gave to utterance $u$ when tested against the same speaker's enrollment. 
Thus, WCU combines individual phonetic contributions to estimate the speaker match score for a positive trial.  
We create a set of 3,150 positive enrollment-test pairs on which to fit these weights using the TIMIT dataset~\cite{garofolo1993timit}.
We extract $\boldsymbol{p_u}$ vectors for each of the test utterances as described above, use them to fit the regression model, and fix the weights of the WCU.
These weights are used for all the experiments in the paper.  

We use the Quartznet automatic speech recognition (ASR) system~\cite{kriman2019quartznet} to obtain the transcripts of the test utterances. 
Quartznet is a 1D time-separable convolutional neural network architecture trained on both telephonic conversational and read-speech wideband datasets. 
The character-level transcription is obtained by performing n-best hypothesis using a beam width of 50. 
We convert the transcription to a sequence of phonemes using the grapheme to phoneme converter~\cite{g2pE2019}.
Finally, the set of unique phonemes in the utterance form the phoneme presence indicator vector $\boldsymbol{p_u}$ and the CU and WCU measures are computed using Eq. 1 and 2.

\section{Datasets}
\label{sec:datasets}

\begin{table}[tb]
\small
\centering
\caption{Test utterance net speech and phonetic richness across evaluation protocols ( mean (std) ).}
\begin{tabular}{ccc}
\hline
Protocol          & Net Speech (s) & CU         \\ \hline
voxceleb1         & 7.3 (5.3)      & 26.4 (5.0) \\
voxceleb1.5s      & 4.3 (0.4)      & 23.2 (3.4) \\
voxceleb1.2s      & 1.8 (0.2)      & 13.9 (3.8) \\
voxceleb1.1s      & 0.8 (0.2)      & 7.6 (3.7)  \\
aplawd            & 0.6 (0.1)      & 3.0 (0.8)  \\
aplawd-repetitive & 3.6 (2.0)      & 9.0 (5.0) \\
\hline
\end{tabular}
\label{tab:probe_stats}
\end{table}
 
The \textbf{Voxceleb1-E} \cite{Nagrani17} protocol is a popular speaker verification  benchmark consisting of 40 speakers, 4,715 speaker models, and 4,713 test utterances. 
The official protocol consists of 18,860 positive and negative model-test trials each.
In addition, several short utterance versions of this protocol are created following the method in \cite{kye2020metalearning}: a \textbf{5s-}, \textbf{2s-}, and \textbf{1s- Voxceleb1-E} protocol are each made by taking random clips of the target size of each test probe. 
Probes are repeated to reach sufficient duration if needed prior to random clipping. 
We discard any clips with content that is unintelligible by our ASR model.
The enrollment audios are left unchanged.

Next we create a protocol using the \textbf{Aplawd} dataset \cite{aplawd}. The Aplawd dataset consists of 10 subjects--5 males and 5 females. The subjects provide speech samples for 5 phonetically diverse sentences, 10 digits, and 66 isolated words, all repeated 10 times\footnote{Some subjects do not supply recordings for all of the transcripts.}.
Each speaker's sentences are reserved for forming a single high-quality enrollment, leading to a mean model net speech of 117.8s ($\sigma$=15.4s) and a constant CU of 39 (the maximum value). 
Meanwhile, the short one-word utterances are used for testing. 
The net-speech of the short test utterances is around 0.6 seconds on average and the CU is around 3.
Matching gender trials are created resulting in 5,107 positive and 20,428 negative trials. 
Notably, this dataset has much cleaner audio compared to Voxceleb1 and can better assess phonetic richness's utility in the absence of noise.

In order to test the impact of utterances with low phonetic variability but higher net-speech, we create a test dataset and evaluation protocol containing repeated words.
This protocol uses the Aplawd dataset, and thus we name it \textbf{Aplawd-Repetitive}. Again, we designate the spoken sentences for enrolling the speakers. Test utterances are constructed by concatenating the single word audios so that net speech and phonetic richness can be independently controlled by varying the utterance length in words and the number of unique words to use respectively.
Each probe consists of between 2 and 10 concatenated single-word utterances and between 1 and 10 unique words. The number of total words and unique words are chosen independently for each probe.
The resulting probes contain an average of 2.5 repeated words per probe.
Note that we do not concatenate digital copies of single-word utterances, but instead make use of the repeated utterances in the Aplawd dataset.  
This protocol contains 1,600 positive and 6,400 negative trials.
See table~\ref{tab:probe_stats} for the details of all our evaluation protocols\footnote{The Aplawd and Aplawd-Repetitive protocols are publicly available at \url{https://doi.org/10.5281/zenodo.11663092}}.

\section{System Description}
\label{sec:system}
\subsection{Speaker Verification System}
We perform speaker verification experiments using the Emphasized Channel Attention Propagation and Aggregation Time-Delay Neural Network (ECAPA-TDNN) architecture~\cite{desplanquesTD20}. 
The ECAPA-TDNN architecture which improves the TDNN architecture with multi-headed attention and squeeze excitations has achieved state-of-the-art performance on several speaker verification benchmarks. 
We use the pretrained model checkpoint\footnote{\url{https://huggingface.co/speechbrain/spkrec-ecapa-voxceleb}} trained on the Voxceleb1~\cite{Nagrani17} + Voxceleb2~\cite{chung2018voxceleb2} datasets provided in the SpeechBrain toolkit~\cite{ravanelli2021speechbrain} for all of our experiments.
At inference time, the output of the final fully connected layer is used as the 192-dimensional speaker embedding.
The voice match scoring is performed using cosine distance between the enrollment and test embeddings.
The cosine distance scores are further calibrated using net-speech and phonetic richness to compute the final score. 

\subsection{Score Calibration}
We assess our phonetic richness features, net speech, and their combinations in their ability to add value in calibrating the authorization score. 
We fit a logistic regression (LR) model~\cite{pigeon2000applying,brummer2007fusion} on the combination of raw scores with net-speech, CU, and WCU to classify matched from mismatched trials.
We perform stratified 5-fold cross validation with class weighting to fit the LR model and compute the results on the test sets.
Finally, the performance of the resulting calibrated scores is reported in terms of equal error rate (EER) as well as $minC_{primary}$, a performance metric defined by~\cite{sadjadi2020nist} as an average of the detection cost function at two operating points.

\section{Experiments \& Results}
\label{sec:eval}
\subsection{Correlation Analysis}

We assess our phonetic richness measures and net-speech in terms of their correlation with the ASV score, utilizing Kendall's $\tau$ correlation coefficient~\cite{abdi2007kendall}. 
Additionally, we visualize these relationships for the Aplawd-Repetitive protocol since it is constructed to have lower correlation between net speech and phonetic richness.

\begin{table}[tb]
\centering
\small
\caption{Kendall's $\tau$ measuring the correlation of phonetic richness and net-speech with the ASV scores for the positive examples in each evaluation protocol.}
\begin{tabular}{cccc}
\hline
Protocol          & CU    & WCU   & Log NS \\ \hline
voxceleb1         & 0.148 & 0.142 & 0.166 \\
voxceleb1\_5s     & 0.033 & 0.028 & 0.032  \\
voxceleb1\_2s     & 0.087 & 0.090 & 0.066  \\
voxceleb1\_1s     & 0.168 & 0.163 & 0.153  \\
aplawd            & 0.201 & 0.197 & 0.202  \\
aplawd-repetitive & 0.633 & 0.608 & 0.348  \\
\hline
\end{tabular}
\label{tab:kendall_tau}
\end{table}

\begin{figure}[!tb]
    \centering
   
    \includegraphics[width=\linewidth]{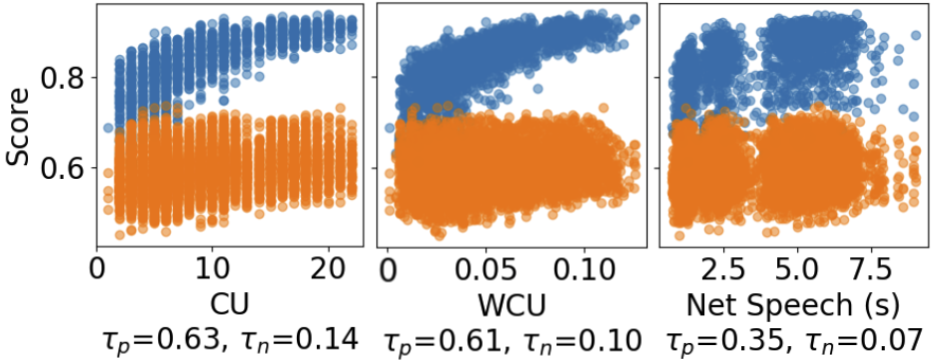}
    \caption{\small ASV score as a function of phonetic richness measures and net speech for \textit{Aplawd-Repetitive} test utterances. Positive (blue) and negative (orange) pairs are plotted separately to observe class separation patterns, and Kendall's $\tau$ is computed for each class.}
    \label{fig:aplawd_rep_corr} 
\end{figure}

In table~\ref{tab:kendall_tau}, we observe that the correlation between ASV score and our phonetic richness measures is similar to net-speech except on the Aplawd-Repetitive protocol. 
By design, the examples in Aplawd-Repetitive have less correlation between phonetic richness and net speech and therefore the independent effects of each on ASV score are easier to distinguish. 
We observe that phonetic richness is significantly more correlated with ASV score than net speech, indicating that it is a more accurate measure of audio quality when it comes to speech where there is repetitive content as shown in figure ~\ref{fig:aplawd_rep_corr}.
In figure~\ref{fig:aplawd_rep_corr}, we additionally observe that the separation between the positive and negative trials is better for utterances with high phonetic richness compared to high net-speech.

\begin{table*}[t]
\centering
\small
\caption{Results of speaker verification score calibration with different feature sets (EER / $minC_{primary}$).}
\begin{tabular}{ccccccc}
\hline
Calibration Features & voxceleb1 & voxceleb1\_5s & voxceleb1\_2s & voxceleb1\_1s & aplawd-repetitive & aplawd \\ \hline
None      & 1.04 / 0.125 & 1.15 / 0.144 & 2.24 / 0.232 & 5.68 / 0.495 & 1.94 / 0.074 & 5.21 / 0.403 \\ \hline
LNS       & 0.99 / 0.110 & 1.16 / 0.146 & 2.22 / 0.233 & 5.58 / 0.500 & 1.62 / 0.082 & 4.97 / 0.395 \\ \hline
CU        & 1.00 / 0.113 & 1.16 / 0.147 & 2.18 / 0.233 & 5.64 / 0.502 & 1.17 / 0.064 & 4.85 / 0.411 \\
WCU       & 1.01 / 0.114 & 1.16 / 0.146 & 2.20 / 0.241 & 5.62 / 0.505 & 1.24 / 0.093 & 5.03 / 0.445 \\ \hline
LNS + CU  & 0.98 / 0.109 & 1.17 / 0.146 & 2.19 / 0.235 & 5.48 / 0.506 & 1.12 / 0.067 & 4.70 / 0.404 \\
LNS + WCU & 0.98 / 0.110 & 1.17 / 0.145 & 2.21 / 0.242 & 5.50 / 0.506 & 1.25 / 0.094 & 4.74 / 0.423 \\ \hline
\end{tabular}
\label{tab:score_calibration}
\end{table*}

\begin{figure*}[!tb]
    \centering
    \includegraphics[width=\linewidth]{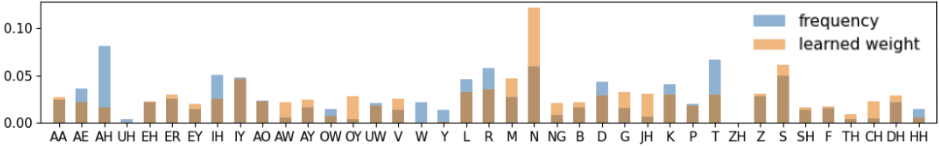}
    \caption{\small Distribution of learned phoneme-specific weights compared with the frequency that each phoneme occurs in the data used for fitting the weights. Phonemes having a larger weight relative to their frequency suggests that they are more useful for carrying speaker-identifying information.}
    \label{fig:weights_hist} 
\end{figure*}

\subsection{Phonetic Richness Based Score Calibration}

Table~\ref{tab:score_calibration} summarizes the results of score calibration using information about the test utterances' phonetic richness and net speech. 
We observe that score calibration with phonetic richness improves the EER on all but one of the benchmarks.
Consistent with the findings of our correlation analysis, phonetic richness is significantly more useful for score calibration than net speech on aplawd-repetitive which contains examples of repetitive speech. 
The EER on Voxceleb1 evaluation reduces from 1.04\% to 1.00\% with score calibration using CU.
When CU is combined with log net-speech (LNS), the EER reduces further to 0.98\%. 
Notably, the combination of net speech and phonetic richness features outperforms net speech when used alone on all protocols except Voxceleb1-2s. 
Finally, we see that as the utterance length of  Voxceleb1 is decreased (see Voxceleb1-2s and Voxceleb1-1s results), the benefit that phonetic richness provides over no-calibration increases and the benefit of adding phonetic richness over net speech alone increases, indicating that phonetic richness provides greater benefit for short utterance verification.

\subsection{Analysis of Learned Phoneme-Weights}

The WCU-measure models the speaker verification score of same-speaker enrollment-test utterance pairs as an additive model, where each phoneme in a given test utterance contributes a non-negative component to the resulting verification score. 
Thus the learned phoneme-specific weights can be interpreted to give insights into the importance of different phonemes for carrying speaker-identifying information. 
To facilitate this analysis, the learned weights are normalized to sum to one and compared against the train dataset's corresponding phoneme frequencies in figure~\ref{fig:weights_hist}. 
The normalized weights are often similar to the phoneme frequencies. 
Cases where the weight differs greatly from the frequency may reflect on the phoneme's ability to provide speaker information. 
A previous work that studied the utility of different syllable categories in identifying speakers found that the nasals (`M', `N', `NG') and affricates (`ZH', `CH', `SH', `R', `J', `Q') combined with the vowels are the most useful for carrying speaker identifying information \cite{6376657}.
We observe that the learned weights for each of the nasal consonant phonemes along with the affricates `CH' and `JH' are all significantly higher than their corresponding phoneme frequencies. 
Furthermore, the `N' and `M' phonemes received the top 1 and 3 largest weights respectively. 
Thus, these results support the previous findings in the literature.

\section{Conclusions}
\label{sec:concl}
In this paper, we investigate phonetic richness as a measure of utterance quality for speaker verification.
We define the count unique (CU) and phoneme-weighted count unique (WCU) measures which are positively correlated ($\tau=0.6$) with the cosine scores of the speaker verification system for positive examples in the short utterance Aplawd-Repetitive dataset.
Phonetic richness is more correlated with the speaker-match score for short utterances.
The proposed CU and WCU measures are useful for score calibration on the Voxceleb-1 benchmark as well as the Aplawd dataset.
Based on the results, we conclude that phonetic richness is more helpful than net speech for calibration when speech contains repetitive content.
Phonetic richness and net speech are complimentary, yielding improved performance over net speech alone on all but one of our evaluation protocols.
The phoneme weighted phonetic richness measure for calibration is not found to be better on short utterances than the non-weighted measure.
In the future, we plan to explore other measures of phonetic richness without explicitly performing ASR transcription.
We also plan to explore calibration based on the phonetic richness of the enrollments. 
Further study is needed to explore the efficacy of this method on multiple languages and cross-lingual speaker verification.

\bibliographystyle{IEEEtran}
\bibliography{refs}

\begin{thebibliography}{10}
\providecommand{\url}[1]{#1}
\csname url@samestyle\endcsname
\providecommand{\newblock}{\relax}
\providecommand{\bibinfo}[2]{#2}
\providecommand{\BIBentrySTDinterwordspacing}{\spaceskip=0pt\relax}
\providecommand{\BIBentryALTinterwordstretchfactor}{4}
\providecommand{\BIBentryALTinterwordspacing}{\spaceskip=\fontdimen2\font plus
\BIBentryALTinterwordstretchfactor\fontdimen3\font minus \fontdimen4\font\relax}
\providecommand{\BIBforeignlanguage}[2]{{%
\expandafter\ifx\csname l@#1\endcsname\relax
\typeout{** WARNING: IEEEtran.bst: No hyphenation pattern has been}%
\typeout{** loaded for the language `#1'. Using the pattern for}%
\typeout{** the default language instead.}%
\else
\language=\csname l@#1\endcsname
\fi
#2}}
\providecommand{\BIBdecl}{\relax}
\BIBdecl

\bibitem{Zeinali2019}
\BIBentryALTinterwordspacing
H.~Zeinali, K.~A. Lee, J.~Alam, and L.~Burget, ``Short-duration speaker verification (sdsv) challenge 2021: the challenge evaluation plan,'' 12 2019. [Online]. Available: \url{https://arxiv.org/abs/1912.06311v3}
\BIBentrySTDinterwordspacing

\bibitem{mandasari2015quality}
M.~I. Mandasari, R.~Saeidi, and D.~A. Van~Leeuwen, ``Quality measures based calibration with duration and noise dependency for speaker recognition,'' \emph{Speech Communication}, vol.~72, pp. 126--137, 2015.

\bibitem{Thienpondt2020TheIV}
\BIBentryALTinterwordspacing
J.~Thienpondt, B.~Desplanques, and K.~Demuynck, ``The idlab voxsrc-20 submission: Large margin fine-tuning and quality-aware score calibration in dnn based speaker verification,'' \emph{ICASSP 2021 - 2021 IEEE International Conference on Acoustics, Speech and Signal Processing (ICASSP)}, pp. 5814--5818, 2020. [Online]. Available: \url{https://api.semanticscholar.org/CorpusID:225041173}
\BIBentrySTDinterwordspacing

\bibitem{karam2011towards}
Z.~N. Karam, W.~M. Campbell, and N.~Dehak, ``Towards reduced false-alarms using cohorts,'' in \emph{2011 IEEE international conference on acoustics, speech and signal processing (ICASSP)}.\hskip 1em plus 0.5em minus 0.4em\relax IEEE, 2011, pp. 4512--4515.

\bibitem{Tan2018DNNBasedSC}
\BIBentryALTinterwordspacing
Z.~Tan, M.~wai Mak, and B.~K.-W. Mak, ``Dnn-based score calibration with multitask learning for noise robust speaker verification,'' \emph{IEEE/ACM Transactions on Audio, Speech, and Language Processing}, vol.~26, pp. 700--712, 2018. [Online]. Available: \url{https://api.semanticscholar.org/CorpusID:3402667}
\BIBentrySTDinterwordspacing

\bibitem{rao2021improving}
H.~Rao, K.~Phatak, and E.~Khoury, ``Improving speaker recognition with quality indicators,'' in \emph{2021 IEEE Spoken Language Technology Workshop (SLT)}.\hskip 1em plus 0.5em minus 0.4em\relax IEEE, 2021, pp. 338--343.

\bibitem{Lavrentyeva2022InvestigationOD}
\BIBentryALTinterwordspacing
G.~Lavrentyeva, S.~Novoselov, A.~Shulipa, M.~Volkova, and A.~Kozlov, ``Investigation of different calibration methods for deep speaker embedding based verification systems,'' \emph{ArXiv}, vol. abs/2203.15106, 2022. [Online]. Available: \url{https://api.semanticscholar.org/CorpusID:247778556}
\BIBentrySTDinterwordspacing

\bibitem{Cumani2021AGM}
\BIBentryALTinterwordspacing
S.~Cumani and S.~Sarni, ``A generative model for duration-dependent score calibration,'' in \emph{Interspeech}, 2021. [Online]. Available: \url{https://api.semanticscholar.org/CorpusID:239711242}
\BIBentrySTDinterwordspacing

\bibitem{Mandasari2013QualityMF}
\BIBentryALTinterwordspacing
M.~I. Mandasari, R.~Saeidi, M.~McLaren, and D.~A. van Leeuwen, ``Quality measure functions for calibration of speaker recognition systems in various duration conditions,'' \emph{IEEE Transactions on Audio, Speech, and Language Processing}, vol.~21, pp. 2425--2438, 2013. [Online]. Available: \url{https://api.semanticscholar.org/CorpusID:13979425}
\BIBentrySTDinterwordspacing

\bibitem{6639154}
T.~Hasan, R.~Saeidi, J.~H.~L. Hansen, and D.~A. van Leeuwen, ``Duration mismatch compensation for i-vector based speaker recognition systems,'' in \emph{2013 IEEE International Conference on Acoustics, Speech and Signal Processing}, 2013, pp. 7663--7667.

\bibitem{Poddar2017SpeakerVW}
\BIBentryALTinterwordspacing
A.~Poddar, M.~Sahidullah, and G.~Saha, ``Speaker verification with short utterances: a review of challenges, trends and opportunities,'' \emph{IET Biom.}, vol.~7, pp. 91--101, 2017. [Online]. Available: \url{https://api.semanticscholar.org/CorpusID:3923424}
\BIBentrySTDinterwordspacing

\bibitem{larcher:hal-01927726}
\BIBentryALTinterwordspacing
A.~Larcher, K.~A. Lee, B.~Ma, and H.~Li, ``{The RSR2015: Database for Text-Dependent Speaker Verification using Multiple Pass-Phrases},'' in \emph{{Annual Conference of the International Speech Communication association (Interspeech)}}, Portland, United States, Sep. 2012. [Online]. Available: \url{https://hal.science/hal-01927726}
\BIBentrySTDinterwordspacing

\bibitem{KANAGASUNDARAM201469}
\BIBentryALTinterwordspacing
A.~Kanagasundaram, D.~Dean, S.~Sridharan, J.~Gonzalez-Dominguez, J.~Gonzalez-Rodriguez, and D.~Ramos, ``Improving short utterance i-vector speaker verification using utterance variance modelling and compensation techniques,'' \emph{Speech Communication}, vol.~59, pp. 69--82, 2014. [Online]. Available: \url{https://www.sciencedirect.com/science/article/pii/S0167639314000053}
\BIBentrySTDinterwordspacing

\bibitem{yolchuyeva2019grapheme}
S.~Yolchuyeva, G.~N{\'e}meth, and B.~Gyires-T{\'o}th, ``Grapheme-to-phoneme conversion with convolutional neural networks,'' \emph{Applied Sciences}, vol.~9, no.~6, p. 1143, 2019.

\bibitem{g2pE2019}
K.~Park and J.~Kim, ``g2pe,'' \url{https://github.com/Kyubyong/g2p}, 2019.

\bibitem{garofolo1993timit}
J.~S. Garofolo, ``Timit acoustic phonetic continuous speech corpus,'' \emph{Linguistic Data Consortium, 1993}, 1993.

\bibitem{kriman2019quartznet}
S.~Kriman, S.~Beliaev, B.~Ginsburg, J.~Huang, O.~Kuchaiev, V.~Lavrukhin, R.~Leary, J.~Li, and Y.~Zhang, ``Quartznet: Deep automatic speech recognition with 1d time-channel separable convolutions,'' 2019.

\bibitem{Nagrani17}
A.~Nagrani, J.~S. Chung, and A.~Zisserman, ``Voxceleb: a large-scale speaker identification dataset,'' in \emph{INTERSPEECH}, 2017.

\bibitem{kye2020metalearning}
S.~M. Kye, Y.~Jung, H.~B. Lee, S.~J. Hwang, and H.~Kim, ``Meta-learning for short utterance speaker recognition with imbalance length pairs,'' 2020.

\bibitem{aplawd}
R.~Serwy, ``Aplawd markings database,'' \url{https://github.com/serwy/aplawdw}, 2017.

\bibitem{desplanquesTD20}
B.~Desplanques, J.~Thienpondt, and K.~Demuynck, ``{ECAPA-TDNN:} emphasized channel attention, propagation and aggregation in {TDNN} based speaker verification,'' in \emph{Interspeech 2020}, H.~Meng, B.~Xu, and T.~Fang~Zheng, Eds.\hskip 1em plus 0.5em minus 0.4em\relax {ISCA}, 2020, pp. 3830--3834.

\bibitem{chung2018voxceleb2}
J.~S. Chung, A.~Nagrani, and A.~Zisserman, ``Voxceleb2: Deep speaker recognition,'' \emph{arXiv preprint arXiv:1806.05622}, 2018.

\bibitem{ravanelli2021speechbrain}
M.~Ravanelli, T.~Parcollet, P.~Plantinga, A.~Rouhe, S.~Cornell, L.~Lugosch, C.~Subakan, N.~Dawalatabad, A.~Heba, J.~Zhong \emph{et~al.}, ``Speechbrain: A general-purpose speech toolkit,'' \emph{arXiv preprint arXiv:2106.04624}, 2021.

\bibitem{pigeon2000applying}
S.~Pigeon, P.~Druyts, and P.~Verlinde, ``Applying logistic regression to the fusion of the nist'99 1-speaker submissions,'' \emph{Digital Signal Processing}, vol.~10, no. 1-3, pp. 237--248, 2000.

\bibitem{brummer2007fusion}
N.~Brummer, L.~Burget, J.~Cernocky, O.~Glembek, F.~Grezl, M.~Karafiat, D.~A. Van~Leeuwen, P.~Matejka, P.~Schwarz, and A.~Strasheim, ``Fusion of heterogeneous speaker recognition systems in the stbu submission for the nist speaker recognition evaluation 2006,'' \emph{IEEE Transactions on Audio, Speech, and Language Processing}, vol.~15, no.~7, pp. 2072--2084, 2007.

\bibitem{sadjadi2020nist}
S.~O. Sadjadi, C.~S. Greenberg, E.~Singer, D.~A. Reynolds, and L.~Mason, ``Nist 2020 cts speaker recognition challenge evaluation plan,'' 2020.

\bibitem{abdi2007kendall}
H.~Abdi, ``The kendall rank correlation coefficient,'' \emph{Encyclopedia of measurement and statistics}, vol.~2, pp. 508--510, 2007.

\bibitem{6376657}
N.~Fatima and T.~F. Zheng, ``Syllable category based short utterance speaker recognition,'' in \emph{2012 International Conference on Audio, Language and Image Processing}, 2012, pp. 436--441.

\end{thebibliography}

\end{document}